\documentclass[10pt, conference, letterpaper]{IEEEtran}

\usepackage{graphicx}
\usepackage[english]{babel}
\usepackage[compatibility=false]{caption}
\usepackage{booktabs}
\usepackage{caption}
\usepackage{subcaption}
\usepackage{setspace}
\usepackage{amssymb}
\usepackage[outdir=./]{epstopdf}
\usepackage{multirow}
\usepackage{verbatim}
\usepackage{amsfonts}
\usepackage{comment}
\usepackage{color}
\usepackage{amsmath}
\usepackage[utf8]{inputenc}
\usepackage{algorithm2e}
\usepackage{algorithmic}
\usepackage{stfloats}

\usepackage[numbers,sort&compress]{natbib}
\usepackage{flushend}
\begin{document}
\RestyleAlgo{boxruled}
\LinesNumbered
\sloppy
\title{Reversing The Meaning of Node Connectivity for Content Placement in  Networks of Caches} 

\author{\IEEEauthorblockN{Junaid Ahmed Khan\IEEEauthorrefmark{1}, Cedric Westphal\IEEEauthorrefmark{2}, J.J. Garcia-Luna-Aceves\IEEEauthorrefmark{3}, and Yacine Ghamri-Doudane\IEEEauthorrefmark{4}\\}
\IEEEauthorrefmark{1}New York University, NY, USA \\
\IEEEauthorblockA{\IEEEauthorrefmark{2}Futurewei Technologies, Santa Clara, CA, USA\\
\IEEEauthorrefmark{3}University of California, Santa Cruz, CA, USA\\  \IEEEauthorrefmark{4}L3i Lab, University of La Rochelle, France\\}
\IEEEauthorblockA{junaid.khan@nyu.edu, cwestphal@gmail.com, jj@soe.ucsc.edu,  yacine.ghamri@univ-lr.fr\\}}

\maketitle

\begin{abstract}
\label{sec:abs}
It is a widely accepted heuristic in content caching to place the most popular content at the nodes that are the best connected. The other common heuristic is somewhat contradictory, as it places the most popular content at the edge, at the caching nodes nearest the users. We contend that neither policy is best suited for caching content in a network and propose a simple alternative that places the most popular content at the least connected node. Namely, we populate content first at the nodes that have the lowest graph centrality over the network topology. Here, we provide an analytical study of this policy over some simple topologies that are tractable, namely regular grids and trees. Our mathematical results demonstrate that placing popular content at the least connected nodes outperforms the aforementioned alternatives in typical conditions. 
\end{abstract}
%\IEEEpeerreviewmaketitle

\section{Introduction}
\label{sec:intro}

Caching of content in a large distributed caching system has relied on simple heuristics. Optimal cache and content placement may be derived (see for instance~\cite{ioannidis2016adaptive,ioannidis2017jointly,optimal-cache-ccn} but these solutions require knowledge of the overall demand vector and network topology to solve some complex optimization problems. As a consequence, simple heuristics have been relied upon to place content in caches. 

The two main heuristics are to place the most popular content either at the edge (see, for instance~\cite{fayazbakhsh2013IDICN}), or at the most connected nodes (say,~\cite{cache-lessformore12}). The intuition for the former is that the popular content should be placed near the user. It delivers more "bang for the buck," due to the distribution of the popularity of the content request, that tends to drop down quickly after the most popular pieces of content. The intuition for the latter is that placing the most popular content at the best connected nodes ensures that it will be available to others, who are more likely to reach these nodes and thereby find the content. 

We contend that neither heuristic policy is best suited for caching content in a network and present another simple policy that places the most popular content at the nodes with the least centrality. We denote this policy by {\em LCHP}, for {\em Low Centrality High Popularity}. In this paper, we consider the mathematical performance of LCHP on some simple topologies and demonstrate that it offers better performance to the other policies, namely a greedy placement of the most popular content at the edge (similar to~\cite{fayazbakhsh2013IDICN}) and the HCHP policy that places the {\em High Popularity} content at the nodes with {\em High Centrality}.

Our objective is to demonstrate that other simple heuristics should be considered, in Information-Centric Networks in particular, to place the content in the caches. The main contribution of the paper is to prove with analytical results on the grid and the tree that LCHP outperforms greedy and HCHP for content with a Zipf distribution for the popularity with parameter less than 1.

The paper is organized as follows: Section~\ref{sec:related-work} discusses the related work; Section \ref{sec:model} presents the system model in which LCHP is defined and Section \ref{sec:centrality} describes LCHP in detail and presents a novel centrality metric that we call  Cache-Connectivity Centrality (CCC) to allow nodes to compute their  own centralities.
Section \ref{sec:placement} considers the performance of LCHP in an analytical manner on regular lattice and tree topologies. Section \ref{sec:perf-evaluation} presents the results of the analytical evaluation.  Section \ref{sec:conclusions} concludes our paper with some insights into future directions.
 
% We explore different centrality metrics (namely: degree, closeness, betweenness, eigenvector, neighborhood cache size, defined later) so as to assess which centrality metric offers the best performance to guide local caching and traffic offload. 

\section{Related Work}
\label{sec:related-work}

Centrality is a graph metric that is used to identify the most important nodes in a network. There are different ways to compute such a measure: the degree of a node, the closeness, the betweenness, the eigenvector centrality are common ones. 

Finding epidemic nodes that spread diseases~\cite{spreaders} is an application in medical science, but it can be mapped to finding the nodes most able to diffuse content in our context. Similarly, Google's PageRank \cite{pagerank} algorithm ranks the importance of a specific content (here, a web-page) in an Internet search based on the number of web links directed towards it.  Medya et al. \cite{arxiv-feb-2017} characterize centrality as an important network metric and suggests modifications to the network structure to increase the centrality of a set of nodes to improve performance. 

Wei et al.~\cite{survey-socialrouting} provide an extensive survey on different socially-aware routing protocols for Delay Tolerant Networks (DTNs) and investigates key centrality metrics. For example, BubbleRap~\cite{bubblerap} and ML-SOR~\cite{mlsor} use nodes with high centrality score for data dissemination and routing in opportunistic mobile social networks. 
Pantazoupoulos et al. \cite{wons10} defined a variant of typical betweenness centrality named as conditional betweenness centrality (CBC) to place content. The authors considered a subset of nodes with high CBC for content caching. However, no solution is provided on which content should be placed at each node. 
Chai et al. \cite{cache-lessformore12} proposed centrality-based caching algorithm by exploiting the concept of betweenness centrality to improve the caching gain. 
Wang et al.  \cite{icnp13} solve an optimization problem to find where to place the content in order to minimize the cost of delivery. A solution is provide to address how much cache should be allocated at each node. By contrast, our target is to address which nodes should cache which content in the network.
 
Rossi et al. \cite{infocom-nomen12} proposed to size the cache of different nodes as a function of different centrality metrics such as degree, betweenness, closeness, graph, and eccentricity. In support with our findings below, they show little benefit of using centrality to size caches. 
In a similar context, Yufei et al. \cite{iswcs16} use ``control nodes" to cache content  based on betweeness centrality. 
Nguyen and Nakazato~\cite{ijcnc13} proposed a betweenness-centrality-based network coder placement for peer-to-peer (P2P) content distribution. However, they do not consider content placement; they  use random linear network coding at every coding node to assist content delivery. 

Distributed caching in ICN is targeted by different studies~\cite{dist-cache-icn}. Wang et al.~\cite{optimal-cache-ccn} address the distribution of the cache capacity across routers under a constrained total storage budget for the network. They found that network topology and content popularity are two important factors that affect where exactly should cache capacity be placed. Yu et al. \cite{yu2015congestion} looked at pushing content to the edge to anticipate network congestion, while Azimdoost et al. \cite{azimdoost2016fundamental} computed the capacity of an ad-hoc network of caches. As mentioned above, edge caching~\cite{fayazbakhsh2013IDICN} is a popular scheme in ICN. Optimal content placement can be found~\cite{ioannidis2016adaptive,ioannidis2017jointly}, but only approximations with a provable bound are available for a distributed, dynamic context. 

\section{Background and Assumptions}
\label{sec:model}

We consider a caching network such as an ICN/CCN or network of caches. Each node request the content, either from its neighbors, or from an origin server. We assume the content is requested from a neighborhood that is at most $h$-hops away from the requesting node, or from the origin server if not found within the $h$-hop neighborhood. We assume that each node contains a cache that can serve content to other nodes. 

\begin{comment}
\begin{table}
  	\caption{List of Notations}
  	\label{tab:notations}
  	\begin{center}
	  	\scalebox{0.9}{
  		\begin{tabular}{ l|l }
  			\hline
  			Notation & Description\\ \hline
  			$\mathbb{V}=\{v\}$ & Set of nodes \\
  			$E^v(t) $ &Set of edges at time $t$\\
  			$\mathbb{X}=\{x\}$ & Set of content  \\
  			$A_v^x$ & Content $x$ availability at node $v$\\
 			$b_v$& Node buffer size \\
 		$t$ &   time instant  \\
 		$\mathbb{P}=\{p_x\}$ & Content popularity set \\
%$P_{SP}/G_{SP}/L_{SP}$ &Service Provider profit/gain/loss \\
%$P_{MN}/G_{MN}/L_{MN}$ & Node profit/gain/loss \\
%$d^x_v$ & Content delivery rate \\ 
%$r_v/c_v$ & Node $v$ reward/cost per unit content delivery \\
$ S$& Set of nodes in a neighborhood\\
$CCC $& Cache Connectivity Centrality \\
  			\hline
  		\end{tabular}}
  	\end{center}
  \end{table}
\end{comment}
  
\subsection{Connectivity Model}

Formally, we assume a set of user nodes $\mathbb{U}=\{u\}$ can retrieve content from nearby set of content caching nodes $\mathbb{V}=\{v\}$. If a user is unable to retrieve the content locally, it can download it directly from the origin server. 

The connectivity of  nodes is modeled by a graph $G(\mathbb{V}(t),\mathbb{E}^v(t))$, where $\mathbb{V}=\{{v}\}$ is the set of non-leaf nodes and  $\mathbb{U}=\{u\}$ is the set of users as leaf-nodes in the network. We consider a dynamic network topology for nodes to connected to different nodes at different times, where the time instant $t$ integrates the temporal network connectivity. $\mathbb{E}^v(t)=\{e_{jk}(t) \mid v_j, v_k \in \mathbb{V}, j \neq k\}$ is the set of edges $e_{jk}(t)$ modeling the existence of a communication link between nodes $j$ and $k$ at time instant $t$ where such links are stable for a period of time. 

We assume that the nodes in the proximity model are aware of each other within a given neighborhood, the node $v$ can connect to the nodes for up to $h$ hops in the set $S_v$ within an acceptable delay. This is a pre-requisite to enable proximity communication. For instance, beacons carrying the one-hop neighborhood of a node enable each node to construct a two-hop neighborhood $(h=2)$ of their proximity network.
 
\subsection{Caching Model}

We define the catalog of content to be $\mathbb{X}=\{x_j, j = 1,\ldots, N \}$ where $x_j$ is an indivisible content chunk in the network, and $N$ is the (potentially very large) number of pieces of content. In the remaining of the paper, with no loss of generality, we may consider an individual content chunk $x$.

%\subsubsection*{\textbf{Definition 1}} (Content Availability) 
Content chunk $x \in \mathbb{X}$ availability at a node $v$ is $A_v^x \in \{1,0\}$ where $A_v^x=1$ when content is cached at the node. The probability of content $x$ found in local cache $v$ is  $p_{u,x,v}$, then the probability to download from the origin server is ${p_u}_{,x,SP} = 1 - \sum\limits_v {{p_{u,x,v}}} $, where SP represent service provider.
%$A_v^x = \left\{ {\begin{array}{*{20}{c}}
%	{1,}&{{\rm{x\ cached\ at\ v}}}\\
%	{0,}&{{\rm{otherwise.}}}
%	\end{array}} \right.$
	
%Thus, the content availability is zero for all the content $x \in \mathbb{X}$ not cached at the node at a particular location and time-slot.  

%To model the storage requirements for a particular content $x$, we denote $Q_x$ as the number of chunks/pieces cached at a node, assuming the possibility of having multiple content chunks. The file size for the content can be defined as $f_x=Q_x \cdot x $. Thus, the file size of all the content cached at a node $v$ at location $l$ and time slot $\overline{t}$ is given as ${f_v}(l,\overline{t}) = \sum\limits_x {A_v^x (l,\overline{t}) \cdot {f_x}}$, i.e. the sum of file sizes for all the possible content at location $l$ and time slot $\bar t$.

We consider  varying content popularity at different times where at a particular time instant $t$ content chunks  follow a popularity distribution so that  $\mathbb{X}=\{x_1, x_2 ...\}$ have probabilities $P=\{ p_{x_1}$ $p_{x_2}...\}$ of being cached at a node with $p_{x_1}>p_{x_2}...$. Users request content with a Zipf distribution,  ${\rho _{u,x}} = \frac{{{x^{ - \psi }}}}{{\sum\limits_{x' \ne x,x' \in X}^{\left| X \right|} {\frac{1}{{x{'^\psi }}}} }}$, where $\psi$ is the skewness parameter. 

The content popularity set can be shared with the nodes using three approaches: (i) offline method by the content operator as a control message; (ii) local monitoring by the nodes taking into account the number of user interests for the content; (iii) as part of content header, using some signaling mechanism. We assume the nodes use one of these methods to assess the popularity of content.
 
We model a cost associated for user $u$ to retrieve content $x$ as:  
\begin{equation}
\label{eqn:cost}
{c_{u,x}} = \left\{ {\begin{array}{*{20}{c}}

{{c_{u,x,v}} = \sum\limits_{{e_{u,v}} \in {S_u}} {{e_{u,v}},} }\\
{c_{o} \gg {c_{u,x,v}}}
\end{array}} \right.\begin{array}{*{20}{c}}

{A_v^x = 1}\\
{A_u^x = A_v^x = 0}
\end{array}
\end{equation}

where,  $c_{u,x,v}$ is the cost for the node $u$ to retrieve from a local caching node $v$ as the sum to edges to that node in its  $S_u$ neighborhood, $c_{o} $ is the higher cost for $u$ to download the content $x$ from the origin server.

\section{Centrality-based Content Placement}
\label{sec:centrality}

\subsection{Computing Centrality}

Typical centrality measures are: {\em degree} (the number of directly connected nodes), $D{C_v} = \left| {{\Gamma _v}} \right| = \left\{ {v'\left| {{e_{vv'}} \in {\mathbb{E}^v}} \right.} \right\}$, {\em closeness} (the average length of the shortest paths between the node and all other nodes in the graph), $C{C_v} = \frac{1}{{\sum\limits_{v' \in {S_v}} {d(v'v)} }}$; {\em betweenness} (the number of shortest paths between all pairs of nodes in the graph going through a specific node), $B{W_v} = \sum\limits_{{v_s} \ne v \ne {v_t}} {\frac{{{\sigma _{{v_s}{v_t}}}(v)}}{{{\sigma _{{v_s}{v_t}}}}}}$ and {\em eigenvector centrality} (a measure of node influence in the network), etc. There exist several variants of the above centralities adapted to different network topologies.However, some metrics are computationally intensive to calculate, or do not lend themselves to distributed implementation. Further, these centralities only take into account the {\em network connectivity} but not the {\em cache connectivity}.  Thus, in order to allow nodes compute the respective centrality without relying on an infrastructure network, we use here a centrality metric using local (connectivity and caching) metrics. We denote this centrality as follows.

%We present a caching-aware centrality metric which is the first approach that considers the amount of buffer a node can offer to cache content locally. The goal is to first form a local fog where connected devices self-organize and collaborate for caching content to facilitate content delivery to nearby devices where each device is considered a node in a graph. 
%\subsubsection{Degree Centrality}
%\subsubsection{Closeness Centrality}
%\subsubsection{Betweenness Centrality}
%\subsubsection{Eigenvector Centrality}

\subsubsection*{\bf Definition - Cache Connectivity Centrality ($CCC$)}  It is the total amount of cache reachable within the $h$-hop neighborhood $S_v$ for a node $v$. $CC{C_v} = \sum\limits_{\forall v \in {S_v}} {{b_v}} $, where $b_v$ is the buffer size for the node $v$.

For $h=1$, this is related to the node degree. Indeed, if all nodes provide the same amount of caching, it is exactly proportional to the degree. Later, we consider $h=2$ and $h=3$. It is a natural centrality metric, as nodes with higher connectivity will have more neighbors and therefore more connected caches, and therefore a higher $CCC$. 

Using the node degree would be oblivious to the cache capacity. For our purpose, a node is more "central" if it has access to a lot of caching capacity nearby. Being connected to a lot of nodes with little or no cache would yield a high, say, degree centrality, but would not help with content distribution. 

There is little overhead to compute $CCC$ in a distributed manner, as the nodes in the network need to exchange beacons for connectivity purpose. Nodes in the proximity network need to advertise their presence to others. We assume that they insert in this advertisement the information about their cache and that of their neighbors. If nodes exchange their neighbors' information, then the $CCC$ for $h=2$ is easily computed. Nodes can also include their latest estimate of their $CCC$ in that advertisement, again with little overhead. The value of $h$ in the $CCC$ should match the hop limit that we impose to get the content from the proximity network (to maintain low delays and prevent high congestion in the proximity network). 

\section{LCHP-based Content Placement}
\label{sec:placement}

\begin{figure*}
	\centering
	\begin{subfigure}[b]{0.23\textwidth}
		\includegraphics[width=\textwidth]{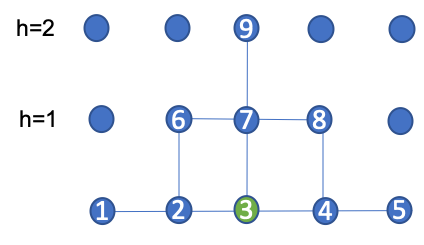}
		\caption{LCHP, h=2}
		\label{fig:lchp_2_proof}
	\end{subfigure}%
	\begin{subfigure}[b]{0.23\textwidth}
		\includegraphics[width=\textwidth]{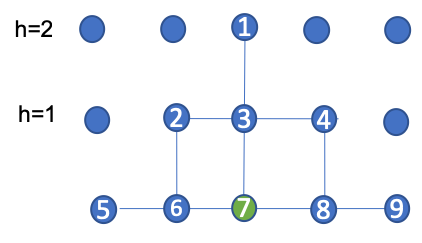}
		\caption{HCHP, h=2}
		\label{fig:hchp_2_proof}
	\end{subfigure}
	\begin{subfigure}[b]{0.23\textwidth}
		\includegraphics[width=\textwidth]{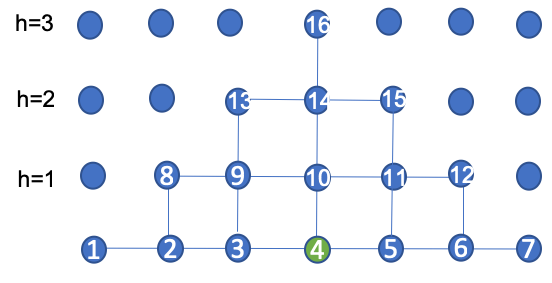}
		\caption{LCHP, h=3}
		\label{fig:lchp_3_proof}
	\end{subfigure}
	\begin{subfigure}[b]{0.23\textwidth}
		\includegraphics[width=\textwidth]{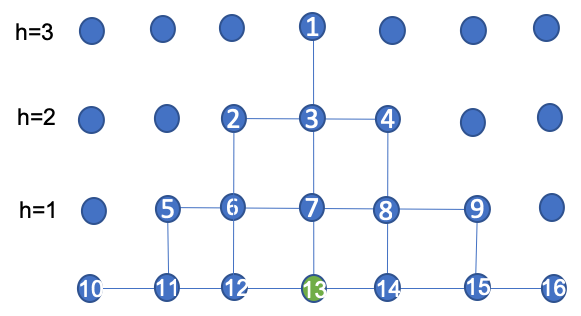}
		\caption{HCHP, h=3}
		\label{fig:hchp_3_proof}
	\end{subfigure}
	\caption{Example grid topology}
	\label{fig:grid_proof}
	% \vspace{-2.5mm}
\end{figure*}

We contend that most popular content should be placed at the nodes with the lowest centrality. We denote the policy that places the High Popularity content at Low Centrality nodes as LCHP. The intuition is that these nodes will have the most difficulty finding the content, due to their poor graph connectivity, and therefore should be given priority for content placement. 

\subsection{Theoretical Analysis on the Lattice} %For a regular grid topology, the content placement LCHP minimizes the above loss function $L$ such that, ${L_{LCHP}} < {L_{HCHP}}$.

It is untractable to prove that LCHP is better than other policies (in particular, "HCHP" which places high popularity content at the high centrality nodes, and "greedy," which greedily fills up the cache with the most popular requests that it receives) on a generic topology. We focus on this section on a lattice topology, and in a tree topology in the next. 

We assume a regular structured $n \times n$ grid topology with set of users/leaf nodes $\mathbb{U}$  requesting for content set  $\mathbb{X}$ from a set of non-leaf caching nodes $ \mathbb{V}$. We also assume the interest can be forwarded up to $h$ hops distance. Namely, we assume that each node is aware of the content cached within a $h$-hop neighborhood, and if the request is a miss in this $h$-hop neighborhood, it is sent to the origin server. 
 
Considering the buffer availability as the centrality measure such as in CCC, the most central node(s) are in the relative center of the grid, we assume each policy begin placing content either at the nodes with the lowest centrality (LCHP) or at the highest centrality nodes in the grid (HCHP). This can be  shown in the Figure~\ref{fig:grid_proof}  where the numbered nodes $1,2,3$.. represent cached content popularity with $p_{x_1}, p_{x_2}, p_{x_3}$ and so on. 

For an arbitrary user node at the leaf/edge of the grid  (green node), the average distance (hops) to retrieve content with popularity $p_{x_1}, p_{x_2},...,$  in its respective $h=2$ hop neighborhood using LCHP placement is given as:
 \begin{multline}
 \label{eqn:lchp_2}
  \frac{6}{5}({p_x}_{_1} + {p_x}_{_2} + {p_x}_{_3} + {p_x}_{_4} + {p_{{x_5}}}) + \frac{5}{3}({p_x}_{_6} + {p_x}_{_7} + {p_x}_{_8}) +\\ 2{p_x}_{_9} + {c_{o}}\sum\limits_{k > 9} {{p_{{x_k}}}}. 
 \end{multline}
 
 This is because the interior nodes have higher centrality (in the CCC sense, but also degree or betweenness) and therefore the most popular content will be spread out at the edge of the grid\footnote{This is different from greedy, where each node at the edge of the grid would cache content $x_1$ since it is the most requested there.}.
 
 Therefore the nearest content will be any of $x_1$ through $x_5$ and the distance to the content will be 0,1 or 2 hops. Averaging over the five possible configurations yield the first term in Eq~(\ref{eqn:lchp_2}). Similar reasoning on the second and third row give the other two terms. 
 
 Similarly, using HCHP based placement, the content retrieval cost is:
  \begin{multline}
   \frac{6}{5}({p_x}_{_5} + {p_x}_{_6} + {p_x}_{_7} + {p_x}_{_8} + {p_{{x_9}}}) + \frac{5}{3}({p_x}_{_2} + {p_x}_{_3} + {p_x}_{_4}) \\+ 2{p_x}_{_1} + {c_{o}}\sum\limits_{k > 9} {{p_{{x_k}}}}.
    \end{multline}

Further, using a greedy approach will place the most popular content  at the edge node (next to the user leaf node), then place the next popular content 1 hop away and the next most popular content 2 hops away from the user node, the average cost to retrieve content is given as ${p_{x_2}} + 2p_{x_3} + {c_{o}}\sum\limits_{j > 3} {{p_{{x_j}}}}$ where, ${c_{o}}$ represents the cost to retrieve the content from the origin server.
 
Now with the delay constraint bounded at $h=3$ hops from a leaf user, the cost to retrieve content placed using LCHP based policy is:
 \begin{multline}
 \label{eqn:lchp_3}
 \frac{{12}}{7}\sum\limits_{i = 1}^7 {{p_{{x_i}}}}  + \frac{{11}}{5}\sum\limits_{j = 8}^{12} {{p_{{x_j}}}}  + \frac{8}{3}({p_x}_{_{13}} + {p_x}_{_{14}} + {p_x}_{_{15}})\\+ 3{p_x}_{_{16}} + {c_{o}}\sum\limits_{k > 16} {{p_{{x_k}}}} ,
   \end{multline}
    while using HCHP, the cost is:
     \begin{multline}
     \frac{{12}}{7}\sum\limits_{i = 10}^{16} {{p_{{x_i}}}}  + \frac{{11}}{5}\sum\limits_{j = 5}^9 {{p_{{x_j}}}}  + \frac{8}{3}({p_x}_{_2} + {p_x}_{_3} + {p_x}_{_4}) \\+ 3{p_x}_{_1} + {c_{o}}\sum\limits_{k > 16} {{p_{{x_k}}}} ,
     \end{multline}
      and using a greedy approach the cost is given as ${p_{{x_2}}} + 2{p_x}_{_3} 
      + 3p_{x_4}+ {c_{o}}\sum\limits_{k > 4} {{p_{{x_k}}}}.$

For a general case, the cost to retrieve content using LCHP  within $h$ hops can be derived as:
\begin{equation}
\sum_{i=1}^{h+1} \mu_{i,h} \sum_{j=m_{i,h}+1}^{m_{i+1,h}} p_j
\label{eqn:lchp_general}
\end{equation}

where $\mu_{i,h}$ is the average distance to the reachable nodes within $h$-hops of the user on the $i$-th row and $m_{i,h}$ is the total number of reachable nodes on the rows below. Namely, 
\begin{eqnarray}
\mu_{1,h} = h(h+1)/(2h+1) \nonumber \\
\mu_{i,h} = 1 + \mu_{i-1,h-1} \nonumber \\
\mu_{h,h} = h
\end{eqnarray}
and 
\begin{eqnarray}
m_{1,h} = 0 \nonumber \\
m_{2,h} = 2h+1 \nonumber \\
m_{i,h} = m_{i-1,h} + 2(h-i)+1 \nonumber \\
m_{h,h} = (2h+1)^2
\end{eqnarray}

The general relation in Eq~(\ref{eqn:lchp_general}) can be used to obtain the special cases of $h=2$ and $h=3$ in Equations~(\ref{eqn:lchp_2}) and~Eq.(\ref{eqn:lchp_3}) respectively.
% ,  $\frac{{h(h + 1)}}{{2h + 1}}\sum\limits_{i = 1}^{2h + 1} {{p_{{x_i}}}}  + \frac{{2h + 1}}{{h + 1}}\sum\limits_{j = 2(h + 1)}^{{{(h + 1)}^2} - 1} {{p_{{x_j}}}}  + h{p_x}_{_{{{(h + 1)}^2}}} + {c_{o}}\sum\limits_{k > {{(h + 1)}^2}} {{p_{{x_k}}}} ,$using HCHP is, $\frac{{h(h + 1)}}{{2h + 1}}\sum\limits_{i = 2h + 1}^{{{(h + 1)}^2}} {{p_{{x_i}}}}  + \frac{{2h + 1}}{{h + 1}}\sum\limits_{j = h}^{2h} {{p_{{x_j}}}}  + h{p_{{x_1}}} + {c_{o}}\sum\limits_{k > {{(h + 1)}^2}} {{p_{{x_k}}}} ,$ and using a greedy content placement policy is given as $\sum\limits_{i = 2}^h {(i - 1){p_{{x_i}}}}  + {c_{o}}\sum\limits_{j > h} {{p_{{x_j}}}}.$  

As the grid size $n$ grows, boundary conditions (i.e. at the corners/edge of the grid) become marginal and can be ignored, and we can just consider the behavior of the policy on some random edge nodes.

Similarly, one can derive a similar equation for HCHP. The coefficient $\mu_{i,h}$ are identical as for LCHP, but the content is ordered in the reverse order, with the least popular content ($p_{(2h+1)^2-(2h+1)}, p_{(2h+1)^2-2h}, ...p_{(2h+1)^2}$) on the bottom row. 

\subsection{Theoretical Analysis on a Regular Tree}

A similar analysis can be performed on a regular tree. We assume here a binary tree with depth $d>>h$, but the same derivation can be provided on the $n$-ary tree. 

For the binary tree, with LCHP, the content at the edge nodes are again the ones with the lowest centrality (by CCC metric, but also degree or betweenness as well). Therefore, the highest popularity content will reside on the bottom row. 

For $h=2$, there are four nodes reachable within $h$ hops, so the caches will contain $x_1$ through $x_4$. The average distance to the content is for LCHP:
\begin{equation}
(p_1+p_2+p_3) + 2p_4 + {c_{o}}\sum\limits_{k > 4} {p_{x_k}}
\end{equation}

For HCHP, it is:
\begin{equation}
(p_2+p_3+p_4) + 2p_1 + {c_{o}}\sum\limits_{k > 4} {p_{x_k}}
\end{equation}

For greedy, it is:
\begin{equation}
p_2+2p_3 + {c_{o}}\sum\limits_{k > 3} {p_{x_k}}
\end{equation}

Since $p_1>p_4$, it is easy to see that LCHP outperforms HCHP in this context. For greedy, it depends on the value of $p_4$ and $c_{o}$, namely the likelihood that a miss at $p_4$ triggers a request to the origin server and the cost of such a request.

For $h=3$, we have:
\begin{eqnarray}
\label{eq:tree-3}
p_1+p_2+2(p_3+p_4+p_5)+3p_6 + {c_{o}}\sum\limits_{k > 6} {p_{x_k}} \mbox{ LCHP}\\
p_5+p_6+2(p_2+p_3+p_4)+3p_1  + {c_{o}}\sum\limits_{k > 6} {p_{x_k}} \mbox{ HCHP}\\
p_2+2p_3+3p_4 + {c_{o}}\sum\limits_{k > 4} {p_{x_k}} \mbox{ Greedy}\\
\end{eqnarray}

Generic cases can be derived as in the grid case as well for $h > 3$ using the regularity of the topology and the monotonous increase in the centrality when going up the tree from a leaf. 

\subsection{Distributed Content Placement}

We present below a heuristic for a set of nodes to opportunistically cache content in a neighborhood.  

Algorithm \ref{algo:caching} is used for content placement by each node. Node $v$ first defines its $h$-hop neighborhood as the set $S_v \in \mathbb{V}$ and  exchanges information regarding its cache size with the nodes in $S_v$ (Lines 4 and 5). Using the cache size information in the neighborhood, it then computes and exchanges its respective centrality $CCC_v$ (Line 9) with the nodes in $S_v$. Each node compares its $CCC$ with the other nodes in their $S_v$ and in case it has the lowest $CCC$, it initializes the placement by caching the most popular content $x$ from the content list $\mathbb{X}$ in order to have access to the most popular content. In Line 7, a node that has filled up its buffer already indicates so, and is not considered as a candidate to be the lowest $CCC$ node any longer. The next lowest $CCC$ node can then fill up its cache by drawing on the next most popular content. 

A popularity tag can be used for each content by the operator where nodes can recognize content based on its tag, i.e. most popular content tagged as popularity level $1$, lesser popular as level $2$ and so on. For ease of explanation, assume that the amount of content tagged with the same label is comparable to the size of one cache, except for the tail of the content that includes all the content that cannot be cached at all. 

Thus, for a node $v$ in a neighborhood $S$ with $k$ nodes having a lower centrality than $CCC_v$, $v$ will cache content with the $(k+1)_{th}$  popularity level content in its placement policy. This content can be cached opportunistically (if the user sees content with this tag, it keeps it) or the cache can be populated with this content by the SP, say during periods of low demand. We denote by $X_{S}$ the content cached in the neighborhood $S$ and by $X_v$ the content cached at node $v$. 

Line 11 of Algorithm~\ref{algo:caching} ensures that the occupied buffer space $b_v$ does not surpass the total available node buffer $ b_v^t$ while $X_S<\mathbb{X}$ allows the node to cache content with decreasing popularity order (Line 12) as long as there exist content to populate its cache until no more content are left to cache. 
The node updates the amount of individually occupied buffer space $b_{v}$ and  $X_{v}$ the set (list) of content it cached (Line 14). The content availability in the neighborhood $S$ is updated as shown in Line 15 and the set of content cached at the node $X_{v}$ are added to the set of contents in the neighborhood $X_{S}$ (Line 16). 

In order to achieve collaborative caching at the set of nodes in the neighborhood, in the increasing order of node centrality $CCC$, the other caching nodes in the neighborhood $S_v$ perform the same steps and cache content in its decreasing popularity order. 
The algorithm converges until either there is no more content left to cache or the corresponding nodes buffers are full as mentioned in Line 11. In case when a node buffer is full, it can exempt itself from the caching process. The nodes with still buffer remaining repeat the process only now considering the content that is not already placed in the neighborhood to add in the unfilled caches.

\begin{algorithm}[t]
\begin{algorithmic}[1]
	\STATE \textbf{INPUT: }   $S$, $\mathbb{X}$,  $P$,  $h$,  $b_v$, $\forall v \in \mathbb{V}$
		\STATE \textbf{OUTPUT: }  $X_{S_v}$, $X_v \in  \mathbb{X}$ 
		\FOR {each node $v$}
		\STATE Define its $h$-hop neighborhood $S_v \in \mathbb{V}$,\label{lst:defineS}
		\STATE Exchange buffer size $b_v $ in $S_v$\label{lst:exchange_cachesize}
	\STATE  Find $CCC_v$ \label{lst:findLC}
		\IF {buffer $b_v$ is full} 
		\STATE set $CCC_v$ to $+ \infty$ \label{lst:full}
		\ENDIF
	\STATE Exchange $CCC_v$, $\forall v$ in $S_v$  \label{lst:exchangeLC}
		\IF {$CCC_v=\mathop {\arg\min }\limits_{v' \in S} (CCC)$, $ v' \neq v $} \label{lst:line:init}
		\WHILE { ($b_v \leq b_v^t$)  or ($X_S < \mathbb{X}$)} %or ($\sum\limits_s {{f_v}}  = \sum\limits_s {b_v^t} $)
		\label{lst:line:while}
		\STATE $b_{v} \gets \mathop {\arg\max }\limits_{x \in \mathbb{X},x \notin {X_S}} (p_x) $ \label{lst:line:getpopular}
			\ENDWHILE
		\STATE Update $b_{v}, X_{v}$	 \label{lst:line:update-buf-profit}	
		\STATE   $A^x_{S} \gets A^x_{S} + A^x_{v}$ \label{lst:line:updateA}
	\STATE   $X_{S} \gets X_{S} \cup X_{v}$ \label{lst:line:update content}
		\ENDIF
		\ENDFOR
		\STATE \textbf{return} $X_v, X_S$
		
\end{algorithmic}
 %\vspace{-2.5mm}
\caption{Content Placement Algorithm at Node $v$}
\label{algo:caching}
\end{algorithm}

\section{Numerical Results}
\label{sec:perf-evaluation}

\subsection{Lattice}

On the lattice topology, for a Zipf content popularity distribution with $\psi=1$, $N=100$ and $h=2$ hops, i.e., $p_{x_1}=0.19, p_{x_2}=0.10, p_{x_3}=0.06, p_{x_4}=0.05 ,p_{x_5}=0.04, p_{x_6}=0.03,$ etc.,  the cost (number of hops) to retrieve content using LCHP is $0.71+ {c_{o}}\sum\limits_{k > 9} {{p_{{x_k}}}}$ while using HCHP, the cost is $0.91+ {c_{o}}\sum\limits_{k > 9} {{p_{{x_k}}}}$, thus, LCHP costs less than HCHP in retrieving content for leaf users. The greedy placement policy cost to retrieve content in this case is $0.22+ {c_{o}}\sum\limits_{j >32} {{p_{{x_j}}}}$, i.e. retrieves all content with popularity lower than $p_{x_3}$ from the origin server. 

Similarly, for a Zipf content popularity distribution with $\psi=1$ and $h=3$ hops, the cost to retrieve content using LCHP can be estimated as $1.22+ {c_{o}}\sum\limits_{k > 16} {{p_{{x_k}}}}$ while using HCHP, the cost can be $1.63+ {c_{o}}\sum\limits_{k > 9} {{p_{{x_k}}}}$, where greedy placement cost is $0.35+ {c_{o}}\sum\limits_{j > 3} {{p_{{x_j}}}}$, however the latter requests content with popularity higher than $p_{x_4}$ from the origin server. 

\begin{figure}[t]
\centering
	\includegraphics[width=0.39\textwidth]{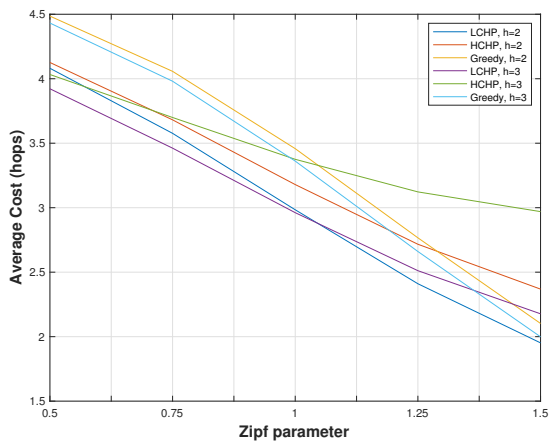}
	\caption{Theoretical comparison of LCHP vs HCHP, grid topology}
	\label{fig:comparison}
%	\vspace{-2mm}
\end{figure}

Figure \ref{fig:comparison} compares LCHP and HCHP for the average cost (hops) for retrieving content by an arbitrary leaf user, when $c_o$ is set to 5 (namely the origin server is 5 hops away; note it has to be at least 3 and 4 hops away for the content to be a miss in the $h$-hop neighborhood when $h=2$ or $h=3$). The cost for different Zipf skewness parameters is shown for $h=2$ and $h=3$. However, irrespective of the variation in the Zipf parameter or the delay bound conditions, LCHP  outperforms HCHP. For $\psi \leq 1$ (which is the typical case for content distribution in the Internet), then LCHP outperforms the other schemes. 

 %\begin{figure}[t]
%\centering
%	\includegraphics[width=0.39\textwidth]{proof.png}
%	\caption{An example regular grid topology}
%	\label{fig:proof}
%	\vspace{-2mm}
%\end{figure}

\subsection{Tree}

 \begin{figure}[t]
\centering
	\includegraphics[width=0.39\textwidth]{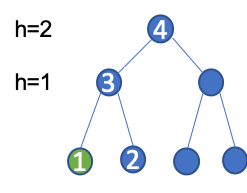}
	\caption{Tree Topology and Placement for LCHP}
	\label{fig:tree-topo}
%	\vspace{-2mm}
\end{figure}

On the tree topology (see Figure~\ref{fig:tree-topo}, for a Zipf content popularity distribution with $\psi=1$ and $h=2$ hops and $N=100$ objects, i.e., $p_{x_1}=0.19, p_{x_2}=0.10, p_{x_3}=0.06, p_{x_4}=0.05$ and $p_{k>4}=0.6$, the cost to retrieve content using LCHP is $0.450+ 0.598{c_{o}}$ while using HCHP, the cost is $0.594+ 0.598{c_{o}}$. Thus, LCHP costs less than HCHP in retrieving content for leaf users. The greedy placement policy cost to retrieve content in this case is $0.22+0.65c_{o}$, i.e. retrieves all content with popularity lower than $p_{x_4}$ from the origin server. This means that for $c_o > 4.6$, LCHP is also better than greedy.

For $h=3$ hops, the cost of LCHP is $0.69+0.35c_o$; the cost of HCHP is $1.07+0.35c_o$; the cost of "greedy" is $0.37+0.60c_o$. The benefit of LCHP is clear versus HCHP, and against greedy if the cost $c_o$ is greater than 1.28. Of course, since the origin server is at least $h$ hops away, $c_o > 3$.

\begin{figure}[t]
\centering
	\includegraphics[width=0.39\textwidth]{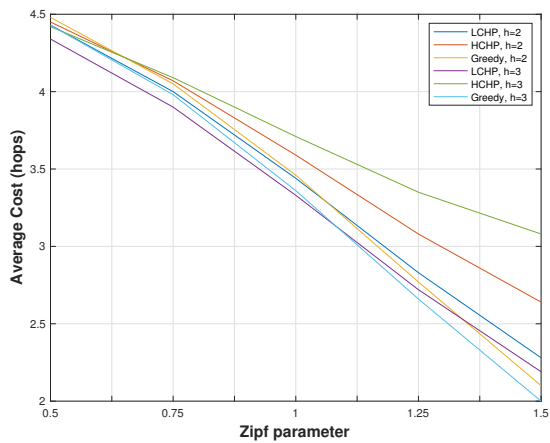}
	\caption{Theoretical comparison of LCHP vs HCHP, tree topology}
	\label{fig:tree-comp}
%	\vspace{-2mm}
\end{figure}

Figure \ref{fig:tree-comp} compares LCHP and HCHP for the average cost (hops) for retrieving content by an arbitrary leaf user in the tree topology. The cost for different Zipf skewness parameters is show for $h=2$ and $h=3$. LCHP always outperforms HCHP. 

\section{Conclusions and Future Work}
\label{sec:conclusions}
 %\vspace{-1mm}

This paper targeted the distributed content placement problem in an information-centric network or a network of caches. We introduce a notion of centrality that takes into account cache size, and propose a centrality-based content placement policy, LCHP. Counter the common intuition of caching content at high centrality nodes or greedily at the edge, LCHP starts placing the more popular content at low centrality nodes since their caching resource needs to be better utilized to compensate for the poor connectivity. We evaluated LCHP along other caching policies using a mathematical analysis on some simple yet illustrative regular topologies: on the grid and the binary tree. LCHP performs the best for Zipf parameter with parameter less than one, which is the common situation in Internet content distribution. This provides some sound basis to further study LCHP on random topologies using simulations, and for designing practical protocols. This is the target of future work. 
%\vspace{-1mm}
\def\bibfont{\footnotesize}

\bibliography{icnc20_final}
\end{document}